# Solution to the Equity Premium Puzzle with Time-Varying Variables

Atilla Aras[a]

[a] Gazi University, Graduate School of Natural and Applied Sciences

Department of Mathematics

Ankara, Türkiye

E-mail: aaras1974@gmail.com

ORCID ID number: 0000-0002-7727-9797

# Solution to the Equity Premium Puzzle with Time-Varying Variables

**Abstract**

The article's aim is to provide a solution to the equity premium puzzle with a derived model. The derived model which depends on Consumption Capital Asset Pricing Model gives a solution to the puzzle with the values of coefficient of relative risk aversion around 4.40 by assuming the subjective time discount factors as 0.97, 0.98 and 0.99. CRRA becomes around 4.11 when the subjective time discount factor is assumed 0.96. These values are found compatible with empirical literature. Moreover, the risk-free asset and equity investors are determined as insufficient risk-loving investors in 1977, which can be considered a type of risk-averse behavior. The risk attitude determination also confirms the validity of the model. Hence, it can be stated that calculated values and risk behavior determination demonstrate the correctness of the derived model because results are robust.

## 1. Background

Mehra and Prescott (1985) and Mehra (2003) demonstrate that some important economic models are unable to give plausible explanations about high risk aversion calculated in standard Consumption Capital Asset Pricing Models (CCAPM), that is, CCAPM cannot have reasonable parameter values in expressing the large historical equity premium. This unreplication is coined equity premium puzzle. Mehra shows this puzzle by possessing price to dividend ratio constant. Aras (2022, 2024) thus provide a solution to the equity premium puzzle presuming this ratio constant. Aras reaches this solution by finding the coefficients of relative risk aversion (CRRA) 1.033526 and 1.058803027 assuming the subjective time discount factor (STDF) 0.99 when testing different data sets. The model in Mehra (2003) because of the constant price to dividend ratio and some other assumptions is not close to the reality in its current form. Hence, this study's aim is to give a solution to the puzzle by possessing the price to dividend ratio time-varying presuming sufficiency factor of the model (SFOM) time-varying.

Presuming time-varying price to dividend ratio enables intertemporal models get one step closer to the reality. Mehra (2003) possesses constant ratio because of the structure of the model. Hence, it can be stated that there is a research gap about possessing solution to equity premium puzzle by making use of time-varying price to dividend ratio in intertemporal models that is inevitable for demonstrating the equity premium puzzle.

The motivation of the study is to have solution for the equity premium puzzle by possessing assumptions that let models be closer to reality. The study's contribution to the existing literature is giving a solution to the equity premium puzzle by having a time-varying ratio and thus possessing working CCAPM under some modifications. Hence, it can be expressed that we have solution for the equity premium puzzle when the price to dividend ratio is in the form of time-varying or constant.

## 2. Literature Review

Scientists agree that STDF is less than 1. It is a coefficient that measures the impatience of investors for consumption. Typical values of this coefficient used in economics and finance are said to be 0.99, 0.98, 0.97, and 0.96.

Many micro literatures suggest that the value of CRRA is about 1. However, some researchers believe that this coefficient may have values of 2, 3, or 4. Despite this, scientists agree that CRRA is less than 10. Levy (2025) expressed that values of CRRA outside the range of 0.75-1.15 gave paradoxical choices. Hence, the writer concluded that CRRA should be close to 1. However, Elmirejad et al. (2025) stated that typical value that has come across in economics area is 1. Additionally, they reported that values in the range of 2-7 were frequently used in finance area.

Large discrepancy between historical real return of equity and real return of risk-free asset cannot be expressed by standard CCAPM under empirically compatible parameters. This is coined equity premium puzzle by Mehra and Prescott (1985). Efforts to provide a solution to this puzzle has been continuing since that date. No agreed solution has emerged yet. Main suggested models are the following.

Rietz (1988) introduced the Rare Disasters hypothesis. The model expresses that investors demand high equity premium for infrequently occurred catastrophic events that lead extreme risks. Barro (2006) then expanded the framework of the model by calibrating disaster probabilities. Epstein and Zin (2013) provided Epstein–Zin (Kreps–Porteus) Preferences that kept apart risk aversion from intertemporal substitution in the recursive utility. Constantinides (1990) and Campbell and Cochrane (1999) provided the habit-formation utility. This utility expresses that an individual utility is equal to current consumption minus habit level.

Constantinides includes external habit formation in consumption-based model. In contrast, Campbell and Cochrane developed consumption surplus ratio. Aiyagari and Gertler (1991) investigated how transaction costs and market imperfection possessed effect on asset returns in their study. Tversky and Kahneman (1992) developed the Cumulative Prospect Theory by refining the Prospect Theory. They add decision weights to cumulative probabilities in their behavioral explanation of equity premium puzzle. Benartzi and Thaler (1995) proposed to combine loss aversion and myopic evaluation concepts from Prospect Theory for equity premium puzzle. Their approach is behavioral. Brown et al. (1995) showed that surviving markets had a positively biased estimate of historical equity returns in their article. Brav et al. (2002) investigated how limited market participation and heterogeneous consumers had effect on asset pricing. Their work supported the hypothesis of incomplete consumption insurance. Ambiguity aversion increases the required rate of return of equity and thereby explains high historical equity premium. For instance, Collard et al. (2018) examined the quantitative effect of ambiguity on asset returns and growth rate in their study.

Over the last five years, efforts to give an agreed solution for equity premium puzzle has continued. Athari et al. (2021) modified the dividend-price ratio and found improved predictability for the equity premium by using US and international data. Barro and Jin (2021) presented a model with rare events and long-run risks. The researchers find the CRRA around 6. Kim (2021) suggested composite asset risk model that combined various risk types for the equity premium puzzle. The writer claims that he explains the puzzle by combining Modern Portfolio Theory with CCAPM. Monache et al. (2021) investigated time-varying relation among price-dividend ratio, expected stock return, and expected dividend growth of US Economy. They find inverse long-run relationship between price-dividend ratio and expected stock return. Shirvani et al. (2021) stated that high equity premium could be expressed if a more suitable distribution was selected for the return data. Aras (2022) introduced the Sufficiency Factor Model that assigned the value of CRRA around 1.03 when STDF was assumed 0.99. He expresses that these values are empirically plausible and confirms the validity of the model. He develops the model by including SFOM to standard CCAPM. Hong (2022) investigated the equity premium puzzle with transaction costs. He finds that CRRA decreases significantly. Chibane (2023) proposed rare disaster models using US consumption data for the equity premium puzzle. The writer expresses that COVID-19 is the hidden disaster risk. Gerotto et al. (2023) presented an agent-based model that replicated the equity premium puzzle. Traders underestimate the risk and overestimate the cost of information in their model. Li and Liu (2023) showed the implications of return extrapolation on CCAPM in their studies. Chung and Fard (2024) handled the puzzle by making use of production-based asset pricing model with GMM. Henry et al. (2024) expressed that financial uncertainty determined the monthly stock market returns. The researchers state that financial uncertainty is stronger predictor than other fourteen macroeconomic predictors. Mai (2024) investigated the main determinants of stock price variations in low and high uncertainty times. They find that cash flow expectations are more important in high uncertainty and recession times. Conversely, discount rates matter more in expansionary times. Inflation expectations are more significant in inflationary environments. Finally, factor returns have a role in determining stock price variations in low and high uncertainty times. Mao et al. (2024) investigated the equity premium puzzle in China and found low CRRA. This result leads them question the existence of the puzzle in China. Aras (2024) confirmed the validity of Sufficiency Factor Model again by determining the theoretically compatible risk attitudes of equity and risk-free asset investors. Lastly, Cassou and Vázquez (2025) explained that choice for consumption predictability gave a solution to the equity premium puzzle in their model.

## 3. Data and Variable Sufficiency Factor Model

### 3.1 Data

We use Mehra and Prescott (1985) data and Table 4 of Mehra (2003). Data involves annual average Standard and Poor's Composite Stock Price Index divided by the consumption deflator, real annual dividends for index series, per capita real consumption on non-durables and services, consumption deflator series, and nominal yield on riskless short-term securities for the 1889-1978 period.

**3.2 Micro Foundation of Sufficiency Factor Model with Time-Varying Price to Dividend Ratio**

Sufficiency Factor Model is the first model in the literature to present that representative agent automatically allocates extra negative or positive utility to uncertain wealth in comparing certain and uncertain utility in the investment process in the market. To put it another way, it is the first model to introduce the hidden variable SFOM.

Investors make comparison between certain and uncertain utility when they invest in the market. They assign extra negative or positive utility to uncertain wealth in this process, and this comparison lets risk attitudes emerge. SFOM is a coefficient that is used for the uncertain utility and takes the form of coefficient when there is future uncertainty and possessing inadequate models for future prediction for representative agents. This coefficient adjusts uncertain utility to make possible to compare it with certain utility. It can be interpreted as follows. SFOM states uncertain utility in terms of certain utility. Investors are then classified as risk-averse, risk-loving and risk-neutral according to the SFOM. It is multiplied by uncertain utility when agents assign extra negative or extra positive utility to uncertain wealth because of the future uncertainty and possessing inadequate models for future prediction.

Mehra (2003) had to presume that price to dividend ratio was constant in the intertemporal models because of the structure of the models. Aras (2022) introduced the Sufficiency Factor Model to give a solution to the equity premium puzzle. Aras (2022, 2024) also assumed the constant ratio by assuming sufficiency factor of the model (SFOM) for equity investors constant. Price to dividend ratio is supposed time-varying by assuming the SFOM for equity investors time-varying in this article.

I formulate the problem of the typical agent to possess the solution for the equity premium puzzle as follows with Time-Varying Sufficiency Factor Model:

$$max_{\{\varsigma_{t+1},\ \varphi_{t+1},\ c_t\}} \{u(c_t) + \psi_m E_t[\sum_{m=t}^{\infty} \beta^{m+1-t} u(c_{m+1})]\}$$

s.t.

$$\varphi_{t+1} s_t + \varsigma_{t+1} p_t + c_t \leq \varsigma_t y_t + \varsigma_t p_t + \varphi_t , \qquad (1)$$

$$c_t \geq 0,\ 0 \leq \varsigma_t \leq 1 \text{ for each t.}$$

The typical agent problem is solved by dynamic programming. This technique alters the problem into two-term periods problem. Investors compare certain and uncertain utility, and their risk attitudes emerge in the market. Hence, this transformation makes it possible to determine the risk behaviors of investors (Aras, 2024). I estimate SFOMs and CRRA from the equation system that is produced from dynamic programming and other methods in the appendix. For instance, year 1977 is present time and year 1978 is future time according to the problem. Hence, the uncertain utility of year 1978 is multiplied by SFOM in two-term period. This process is continuing in the two-term periods until infinity to solve the problem by dynamic programming. Investors compare certain and uncertain utility by including SFOM to uncertain utility and thus make it possible to compare the utilities. Their risk attitudes are specified after the comparison process in each two-terms.

The budget constraint in equation 1 employs the principle of available of resources is equal to use of these resources. Hence, the budget constraint implies that current possession of equity and risk-free asset can be spent as current consumption and next period investment in equity and risk-free asset. Net supply of next period risk-free asset from $s_t$ is zero because maturity of the asset is one year. Budget constraint in equation 1 can be alternatively stated for t+1 as $c_{t+1} \leq \varsigma_{t+1}y_{t+1} + \varsigma_{t+1}p_{t+1} + \varphi_{t+1}R_{f,t+2} - \varphi_{t+2} - \varsigma_{t+2}p_{t+1}$.

We have SFOM for risk-free asset at time t for two reasons:

1-Possessing the uncertainty of $c_{t+1} = \varsigma_{t+1}y_{t+1} + \varsigma_{t+1}p_{t+1} + \varphi_{t+1} - \varphi_{t+2}s_{t+1} - \varsigma_{t+2}p_{t+1}$ at time t.

2- There exists a probability at time t that risk-free asset investors could trade with the FED (The Federal Reserve System) in the open market at a later date during the year. This unclear condition leads to an uncertainty at time t.

The all economy is in equilibrium

1- When we possess market clearing for equity and risk-free asset investors with $\varphi_{t+1}$ = 0 and $\varsigma_t$ = $\varsigma_{t+1}\ldots$ = 1. The first no-trade equilibrium for risk-free assets occurs at time t+1 when we take into account the uncertainty of $c_{t+1}$. Hence, we possess no-trade equilibrium for equity and risk-free asset.

2- When x = z occurs at equilibrium.

We need not assume a specific type of utility curve for the risk-aversity. A risk-averse investor may possess convex or linear utility curve according to new definitions. Despite this, we assume concave utility curve for the typical agent because a risk-averse investor may have a concave utility curve when new definitions are used. Detailed information about this issue can be found at Aras (2023).

We possess the following equations of Sufficiency Factor Model after including the SFOM to standard CCAPM in Mehra (2003) when the SFOM and price to dividend ratio become time-varying. We should also point out that we have to assume price to dividend ratio time-varying provided SFOM for equity investors is presumed time-varying. The assumptions of the Sufficiency Factor Model with time-varying variables are the same as those with constant ones in Aras (2022, 2024).

SOLUTION

$$\ln E(R_{e,t+1}) - \ln E(R_{f,t+1}) = \mu_x(1\text{-}\tau) + 0.5\sigma_x^2\,(1+\tau^2) + \mu_k + 0.5\sigma_k^2 + \rho\sigma_x\sigma_k + \ln\beta + \ln\eta_t\,. \tag{2}$$

See the derivation of equation 2 in the Appendix for the proof.

$$\ln E(R_{f,t+1)}\,(1-\tau\,\rho\sigma_x\sigma_r) - \ln E(R_{e,t+1}) = -\ln\beta\,(\tau\,\rho\sigma_x\sigma_r) + \ln\lambda_t\,(1-\tau\,\rho\sigma_x\sigma_r) - \ln\eta_t(1+\tau\,\rho\sigma_x\sigma_r)\,. \tag{3}$$

See the derivation of equation 3 in the Appendix for the proof.

$$0 = (\ln\beta + \ln\lambda_t)\,[(1-\tau)\,\mu_x + \mu_k + 0.5\sigma_x^2\,(1-\tau)^2 + 0.5\,\sigma_k^2 + (1-\tau)\,\rho\sigma_x\sigma_k]. \tag{4}$$

See the derivation of equation 4 in the Appendix for the proof.

Here,

(1) $u(c,\tau) = \frac{c^{1-\tau}-1}{1-\tau}$;

(2) $\tau$ = coefficient of relative risk aversion;

(3) $c$ is per capita real consumption;

(4) $R_{e,t+1} = \frac{p_{t+1} + y_{t+1}}{p_t}$,

(5) $p_t$ and $y_t$ denote price of the stock and dividend at time t;

(6) $R_{f,t+1} = \frac{1}{s_t}$, where $s_t$ denotes the price of the risk-free asset;

(7) the growth rate of consumption, $x_{t+1} = \frac{c_{t+1}}{c_t}$;

(8) the growth rate of dividends, $z_{t+1} = \frac{y_{t+1}}{y_t}$;

(9) $(x_t\,, z_t)$ are jointly lognormally distributed;

(10) $v_t = \frac{p_t}{y_t}$;

SOLUTION

(11) $k_{t+1} = \frac{v_{t+1}+1}{v_t}$;

(12) E $(R_{e,t+1})$ = mean equity return;

(13) $\mu_x$ = E (ln *x*);

(14) $\mu_k$ = E (ln *k*);

(15) $\sigma_x^2$ = var (ln *x*);

(16) $\sigma_k^2$ = var (ln *k*);

(17) $\beta$ = subjective time discount factor;

(18) $\eta_t$ = sufficiency factor of the model for risk-free asset investors;

(19) $\lambda_t$ = sufficiency factor of the model for equity investors;

(20) $\mu_r$ = E (ln $R_{e,t+1}$);

(21) $\sigma_r^2$ = var (ln $R_{e,t+1}$);

(22) $(x_t, k_t)$ are jointly lognormally distributed;

(23) $(x_t, R_{e,t})$ are jointly lognormally distributed;

(24) $\varsigma_t$ = amount of equity;

(25) $\varphi_t$ = amount of risk-free asset;

(26) $\psi_t$ = sufficiency factor of the model (in the typical agent problem);

(27) E $(R_{f,t+1})$ = mean risk-free rate;

I use equation 1-5 and steps of the new method that exist in Aras (2024) to detect the types of investors of year 1977. I select the year 1977 for risk attitude determination because it is a convergent year (Aras, 2024).

Equations involve SFOM and CRRA. I calculate these variables from the equation 2-4 of this study.

**3.3 Relationships among Variables in the Equation System**

SOLUTION

If we assume all variables constant other than compared variables in the equations, we have the following relationships.

Mean growth rate of consumption increases as gross rate of return of equity increases. Moreover, mean growth rate of consumption decreases as mean growth rate of price to dividend ratio, SFOM for risk-free asset and STDF increase. Lastly, mean growth rate of consumption increases as gross rate of return of risk-free asset decreases.

Gross rate of return of equity increases as SFOM for risk-free asset and STDF increase. Gross rate of return of equity decreases as SFOM for equity increases.

Gross rate of return of risk-free asset increases as STDF decreases. Gross rate of return of risk-free asset decreases as SFOM for risk-free asset increases. Gross rate of return of risk-free asset increases as SFOM for equity increases.

## 4. Results and Discussion

### 4.1 Results

I test the Sufficiency Factor Model with time-varying variables with standard tests (i.e., conditional moments are changed with sample means, variances etc.) I calculate the solution of the system of equations 2-4 by MATLAB. The results are demonstrated in Table 1.

**Table 1**

**Calculation Results of Sufficiency Factor Model with Time Varying Variables in 1977**

| STDF | SFOM (Risk-free Asset) | SFOM (Equity) | CRRA |
|---|---|---|---|
| 0.96 | 1.0928 | 1.0324 | 4.1109 |
| 0.97 | 1.0862 | 1.0232 | 4.3962 |
| 0.98 | 1.0751 | 1.0127 | 4.3968 |
| 0.99 | 1.0642 | 1.0023 | 4.3971 |

SOLUTION

When STDF is varied, its effect on SFOM for risk-free asset and equity investors is not huge. SFOM for risk-free asset investors is around 1.06 to 1.09. Hence, risk-free asset investors allocate extra positive utility for the uncertain wealth. Additionally, SFOM for equity investors is around 1.002 to 1.03. Hence, it can be stated that SFOM is not so sensitive to changes in STDF and equity investors assign extra positive utility to uncertain wealth. When STDF is varied, its effect on CRRA is not large. CRRA remains around 4.40. It becomes around 4.11 when STDF is assumed 0.96. Hence, CRRA is not very much sensitive to varying values of STDF.

Calculated CRRA values in this table are compatible with the existing empirical literature. SFOM for investors are more than 1. Hence it can be said that both risk-free asset and equity investors allocate extra positive utility to uncertain wealth for various values of STDF.

The calculation results of Aras (2022, 2024) are demonstrated in Table 2. SFOM for risk-free asset and equity are assumed constant. This lets price to dividend ratio be constant. STDF is assigned a value of 0.99 for all calculations in this category.

When STDF remains constant at 0.99 value, SFOM for risk-free asset investors is around 1.02. Hence it can be stated that risk-free asset investors assign extra positive utility to uncertain wealth. In contrast, SFOM for equity investors is around 0.96. It means that equity investors extra negative utility to uncertain wealth. When STDF is 0.99, calculated CRRA values are compatible with the existing empirical literature.

SOLUTION

**Table 2**

**Calculation Results of Sufficiency Factor Model with Constant Variables in 1977**

| **Per capita consumption year 1978 (in dollars)** | **STDF** | **SFOM (Risk-free asset)** | **SFOM (Equity)** | **CRRA** |
|---|---|---|---|---|
| 3450 (realized) | 0.99 | 1.019392 | 0.961745 | 1.033526 |
| 3430 (projected) | 0.99 | 1.01997145 | 0.96232455 | 1.058803027 |

Source: Aras (2022, 2024)

When STDF is varied, type of investor of equity investors for year 1977 is insufficient risk-loving, which can be considered a type of risk-averse behavior. This detection is compatible with theory. The results are in Table 3.

When STDF is varied, type of investor of risk-free asset investors for year 1977 is insufficient risk-loving, which can be interpreted as a type of risk-averse behavior. This detection is compatible with theory. The results are shown in Table 4.

Aras (2024) detects the type of investors in year 1977 for equity investors as risk-averse and that of risk-free asset investors as insufficient risk-loving with STDF value of 0.99 and CRRA values of 1.033526 and 1.058803027. This determination is also compatible with theory.

SOLUTION

**Table 3**

**Type of Investors for Equity Investors with Time Varying Variables (realized 1978 value)**

| **STDF** | **CRRA** | **Certain Utility** | **Uncertain Utility** | **Type of investor Year 1977** |
|---|---|---|---|---|
| 0.96 | 4.1109 | 0.32145038 | 0.31797358 | Insufficient risk- loving |
| 0.97 | 4.3963 | 0.29443806 | 0.29223096 | Insufficient risk- loving |
| 0.98 | 4.3968 | 0.29439472 | 0.29214202 | Insufficient risk- loving |
| 0.99 | 4.3971 | 0.29436873 | 0.29209532 | Insufficient risk- loving |

Lastly, it can be stated that both compatible values of STDF and CRRA and the types of investors in year 1977 for risk-free asset and equity investors in this study suggest that Sufficiency Factor Model with time varying variables is correct just like that with constant variables in Aras (2022).

### 4.2 Sufficiency Factor Model and Other Models

On the one hand, habit is added to utility in habit formation models; small persistent growth shocks are applied in long-run risk model; and fear of low probability crashes are included in the utility in rare disaster models. On the other hand, only SFOM is multiplied by uncertain utility in the sufficiency factor models. Mainstream models use recursive or habit utilities. Sufficiency factor models make use of standard CRRA utility.

SOLUTION

**Table 4**

**Type of Investors for Risk-Free Asset Investors with Time Varying Variables (realized 1978 value)**

| **STDF** | **CRRA** | **Certain Utility** | **Uncertain Utility** | **Type of investor Year 1977** |
|---|---|---|---|---|
| 0.96 | 4.1109 | 0.32145038 | 0.07353463 | Insufficient risk- loving |
| 0.97 | 4.3963 | 0.29443806 | 0.06758363 | Insufficient risk- loving |
| 0.98 | 4.3968 | 0.29439472 | 0.06757201 | Insufficient risk- loving |
| 0.99 | 4.3971 | 0.29436873 | 0.06756919 | Insufficient risk- loving |

The Generalized Expected Utility in Epstein and Zin is based on unobserved utility and the proxies they used for unobservables show that they do not reach a solution. The effective risk aversion is impossibly large in habit formation models and this implies the non-existence of solution for the model. Interest rates do not fluctuate as prognosticated in rare disaster models. This shows that these models do not offer solution. Both versions of Sufficiency Factor Model estimate the CRRA smaller than 10. Constant and Variable Sufficiency Factor Model calculates this coefficient about 1.03 and 4.40, respectively. These imply that sufficiency factor models give solution.

Key variables of mainstream models are consumption volatility and rare events. Key variable of sufficiency factor models is the hidden variable SFOM. All mainstream models assume no-trade equilibrium for the assets. However, sufficiency factor models may not assume this for risk-free asset because of the possible trade with FED in the open market. Despite this, sufficiency factor models may assume no-trade equilibrium for all assets.

## 6. Conclusions

This study confirms that derived model is valid. On the one hand, the value of the CRRA remains around 4.40 by presuming the values of the STDF 0.97, 0.98 and 0.99. CRRA becomes around 4.11 when STDF is assumed 0.96. The investors of risk-free asset and equity, on the other hand, are found to be insufficient risk-loving which can be considered a kind of risk-averse behavior. Hence, it can be expressed that risk attitude determination and calculated values are compatible with theory and empirical studies, respectively. All these imply that CCAPM is working under time-varying SFOM.

Derived models, that is both versions of Sufficiency Factor Model with constant and time-varying variables give solution to the equity premium puzzle. Validity of this model is very important for the finance and economics literature because the model is now one step closer to real economic environment by assuming time-varying variables.

Sufficiency factor of the model is not known adequately by researchers. Experimental and behavior economists should study this hidden variable to detect important aspects of it.

Reaching again a solution to the equity premium puzzle with the Sufficiency Factor Model shows the strength of the model. The model provides links between real and financial sectors. Efforts to study on these connections will be very important for policy-making on these sectors because we are now one step closer to the real economic environment with the derived model.

## 7. Statements and Declarations

**Competing Interests:** The author has no relevant financial or non-financial interests to disclose.

**Data Availability Statement:** The data and programs generated (i.e., replication package) for the article are available at https://doi.org/10.7910/DVN/BICAGF, Harvard Dataverse, V4.

**Funding**: No funds, grants, or other support was received.

# Appendix

## Derivation of Equation 2

We possess

$$p_t = v_t y_t \ . \tag{A.1}$$

Then we have

$$R_{e,t+1} = \frac{p_{t+1}+y_{t+1}}{p_t} \ . \tag{A.2}$$

We substitute A.1 in A.2 to possess

$$R_{e,t+1} = \left(\frac{v_{t+1}+1}{v_t}\right)\left(\frac{y_{t+1}}{y_t}\right) . \tag{A.3}$$

We then have

SOLUTION

$$R_{e,t+1} = k_{t+1}\, z_{t+1}. \quad (A.4)$$

If we take conditional expectation of both sides, we will obtain

$$\mathrm{E_t}\,(R_{e,t+1}) = \mathrm{E_t}\,(k_{t+1}\, z_{t+1})\,. \quad (A.5)$$

The equilibrium condition of x = z results in

$$\mathrm{E_t}\,(R_{e,t+1}) = \mathrm{E_t}\,(k_{t+1}\, x_{t+1})\,. \quad (A.6)$$

If we make use of lognormal properties and replace conditional expectations with sample means which is a compatible operation with standard empirical tests, we will obtain

$$\mathrm{E}(R_{e,t+1}) = \exp\,(\mu_x + \mu_k + 0.5\,(\sigma_x^2 + \sigma_k^2 + 2\rho\sigma_x\sigma_k)) \quad (A.7)$$

where $\rho$ = corr (ln *x*, ln *k*).

Take ln of both sides which result in

$$\ln \mathrm{E}(R_{e,t+1}) = \mu_x + \mu_k + 0.5\,(\sigma_x^2 + \sigma_k^2 + 2\rho\sigma_x\sigma_k). \quad (A.8)$$

We have the following from equation A.21 of Aras (2022)

$$R_{f,t+1} = \frac{1}{\beta\eta_t E_t(x_{t+1}^{-\tau})}. \quad (A.9)$$

Take conditional expectation of both sides of A.9 which result in

$$E_t\big(R_{f,t+1}\big) = E_t\Big(\frac{1}{\beta\eta_t E_t(x_{t+1}^{-\tau})}\Big). \quad (A.10)$$

Then A.10 is equal to the following by taking out what is known

$$E_t\big(R_{f,t+1}\big) = \frac{1}{\beta\eta_t E_t(x_{t+1}^{-\tau})}. \quad (A.11)$$

Take ln of both sides of A.11 which result in

$$\ln E_t\big(R_{f,t+1}\big) = \ln 1 - \ln\beta - \ln\eta_t - \ln E_t(x_{t+1}^{-\tau}). \quad (A.12)$$

If we make use of lognormal properties and replace conditional expectations with sample means which is a compatible operation with standard empirical tests, we will obtain

$$\ln E(R_{f,t+1}) = -\ln\beta - \ln\eta_t + \tau\mu_x - 0.5\tau^2\sigma_x^2. \quad (A.13)$$

Subtract A.13 from A.8 to obtain

$\ln E(R_{e,t+1}) - \ln E(R_{f,t+1}) = \mu_x + \mu_k + 0.5\sigma_x^2 + 0.5\sigma_k^2 + \rho\sigma_x\sigma_k + \ln\beta + \ln\eta_t - \tau\mu_x +$

$0.5\tau^2\sigma_x^2$ . (A.14)

We have the following after some algebraic operations

$\ln E(R_{e,t+1}) - \ln E(R_{f,t+1}) = \mu_x(1\text{-}\tau) + 0.5\sigma_x^2\ (1+\tau^2) + \mu_k + 0.5\sigma_k^2 + \rho\sigma_x\sigma_k + \ln\beta +$

$\ln\eta_t$ . (A.15)

**Derivation of Equation 3**

We possess the following from equation A.43 of Aras (2022)

$$\eta_t R_{f,t+1} - \lambda_t E_t(R_{e,t+1}) = \eta_t\lambda_t\ \beta R_{f,t+1}\ cov_t(\frac{u'(c_{t+1})}{u'(c_t)}, R_{e,t+1}) \quad \text{(A.16)}$$

when the sufficiency factor of the model is time-varying.

Because $R_{f,t+1}$ and $E_t(R_{f,t+1})$ are equal when testing with standard tests, we possess

$$\eta_t E_t(R_{f,t+1}) - \lambda_t E_t(R_{e,t+1}) = \eta_t\lambda_t\ \beta E_t(R_{f,t+1}) cov_t(\frac{u'(c_{t+1})}{u'(c_t)}, R_{e,t+1}). \quad \text{(A.17)}$$

We will make use of

$\text{cov}\ (X^a, Y^b) = E(X^a)\ E(Y^b)\ [\exp\ (\text{ab}\rho\sigma_x\sigma_y) - 1]$ where $E(X^a) =$

$\exp\ (\text{a}\mu_x + 0.5\text{a}^2\sigma_x^2)$, $E(Y^b) = \exp\ (\text{b}\mu_y + 0.5\text{b}^2\sigma_y^2)$, $\rho = corr\ (\ln x, \ln y)$ (A.18)

for $cov_t(\frac{u'(c_{t+1})}{u'(c_t)}, R_{e,t+1})$.

Then we possess

$$cov_t(\frac{u'(c_{t+1})}{u'(c_t)}, R_{e,t+1}) =$$

$[\exp\ (-\tau\mu_x + 0.5\tau^2\sigma_x^2)]\ [\exp\ (\ \mu_r + 0.5\sigma_r^2\ )]\ [\exp\ (-\tau\rho\sigma_x\sigma_r) - 1]$, (A.19)

if we make use of lognormal properties and replace conditional covariance with sample covariance which is a compatible operation with standard tests.

Substitute A.19 in A.17 to obtain

$\eta_t E(R_{f,t+1}) - \lambda_t E(R_{e,t+1}) = \eta_t\lambda_t\beta E(R_{f,t+1})[\ \exp\ (-\tau\mu_x + 0.5\tau^2\sigma_x^2 + \mu_r + 0.5\sigma_r^2 -$

$\tau\rho\sigma_x\sigma_r)] - \eta_t\lambda_t\beta E(R_{f,t+1})[\ \exp\ (-\tau\mu_x + 0.5\tau^2\sigma_x^2 + \mu_r + 0.5\sigma_r^2]$ (A.20)

after making use of lognormal properties and substituting conditional expectation with sample average which is a compatible operation with standard tests.

Take ln of both sides to obtain

ln $\eta_t$ + ln $E(R_{f,t+1})$ – ln $\lambda_t$– ln $E(R_{e,t+1})$ = (ln $\beta$ + ln $\eta_t$ + ln $\lambda_t$ + ln $E(R_{f,t+1})$) ($-\tau\mu_x$ + $0.5\tau^2\sigma_x^2$ + $\mu_r$+ $0.5\sigma_r^2$ – $\tau\rho\sigma_x\sigma_r$) – (ln $\eta_t$ + ln $E(R_{f,t+1})$ + ln $\lambda_t$ + ln $\beta$ ) ($-\tau\mu_x$ + $0.5\tau^2\sigma_x^2$ + $\mu_r$+ $0.5\sigma_r^2$). (A.21)

Some algebraic operations after A.21 result in ln $E(R_{f,t+1})$ $(1 - \tau\rho\sigma_x\sigma_r)$ – ln $E(R_{e,t+1})$ = –ln $\beta$ $(\tau\rho\sigma_x\sigma_r)$ + ln $\lambda_t$ $(1 - \tau\rho\sigma_x\sigma_r)$ – ln $\eta_t(1 + \tau\rho\sigma_x\sigma_r)$ . (A.22)

**Derivation of Equation 4**

We possess the following from A.1

$$p_t = v_t y_t \ . \tag{A.23}$$

We also possess the following from equation 18 of Aras (2022)

$$p_t u'(c_t) = \beta\lambda_t \mathrm{E}_t[u'(c_{t+1})(p_{t+1} + y_{t+1})]. \tag{A.24}$$

Some algebraic operations after A.24 and substituting A.23 in A.24 result in

$$v_t y_t = \beta\lambda_t \mathrm{E}_t[(v_{t+1}y_{t+1} + y_{t+1})\frac{u'(c_{t+1})}{u'(c_t)}] \ . \tag{A.25}$$

The above is equal to

$$1 = \beta\lambda_t \mathrm{E}_t(z_{t+1}\, k_{t+1}\, {x_{t+1}}^{-\tau}). \tag{A.26}$$

The equilibrium condition of x = z results in

$$1 = \beta\lambda_t \mathrm{E}_t(k_{t+1}\, {x_{t+1}}^{1-\tau}). \tag{A.27}$$

Then we possess

$$1 = \beta\lambda_t \exp\{(1-\tau)\,\mu_x + \mu_k + 0.5\,[\sigma_x^2\,(1-\tau)^2 + \sigma_k^2 + 2(1-\tau)\,\rho\sigma_x\sigma_k]\} \tag{A.28}$$

where ρ = *corr* (ln *x*, ln *k*) after making use of lognormal properties and substituting conditional expectation with sample average which is a compatible operation with standard tests.

Taking ln of both sides results in

$$0 = (\ln \beta + \ln \lambda_t)\ [(1-\tau)\ \mu_x + \mu_k + 0.5\sigma_x^2\ (1-\tau)^2 + 0.5\sigma_k^2 + (1-\tau)\ \rho\sigma_x\sigma_k]. \tag{A.29}$$